\documentclass[final,3p,times,twocolumn]{elsarticle}

\usepackage{amssymb}

\usepackage{geometry} 
\usepackage{graphicx}  
\usepackage{txfonts}   
\usepackage[colorlinks,linkcolor=blue,anchorcolor=blue,citecolor=blue,urlcolor=blue]{hyperref}
\usepackage{bm}
\usepackage[mathlines]{lineno}
\usepackage{color,soul}

\usepackage{natbib}   
\biboptions{sort&compress}

\journal{Carbon}
\begin{document}


\begin{frontmatter}



\title{{\hfill\small\rm Carbon {\bf 160} (2020)}\\%
       Strain-Controlled Magnetic Ordering in 2D Carbon Metamaterials}



\author[1]{Dan~Liu} %

\author[2]{Eunja~Kim}

\author[3]{Philippe F.~Weck}

\author[1]{David Tom\'{a}nek%
\corref{cor}} %
\ead{tomanek@msu.edu}

\address[1]{Physics and Astronomy Department,
            Michigan State University,
            East Lansing, Michigan 48824, USA}
\address[2]{Department of Physics and Astronomy,
             University of Nevada Las Vegas,
             Las Vegas, NV 89154, USA}
\address[3]{Sandia National Laboratories,
             P.O.~Box 5800,
             Albuquerque, NM 87185, USA}
\cortext[cor]{Corresponding author} %


\begin{abstract}
We use {\em ab initio} spin-polarized density functional theory to
study the magnetic order in a Kagom\'{e}-like 2D metamaterial
consisting of pristine or substitutionally doped phenalenyl
radicals polymerized into a nanoporous, graphene-like structure.
In this and in a larger class of related structures, the
constituent polyaromatic hydrocarbon molecules can be considered
as quantum dots that may carry a net magnetic moment. The
structure of this porous system and the coupling between the
quantum dots may be changed significantly by applying moderate
strain, thus allowing to control the magnetic order and the
underlying electronic structure.
\end{abstract}




\end{frontmatter}


\section{Introduction}

Metamaterials are systems where the separation between the
macrostructure and the constituent material has been washed out to
gain new functionality. 2D metamaterials with selected
nanostructured components can exhibit remarkable mechanical,
transport, optical, and electromagnetic %
properties~\cite{{DT271},{Scalari2012},{Casse2006},{Ishikawa2017}}.
Whereas isolated constituent nanostructures should be considered
as quantum dots, aggregates of such nanostructures may display
intriguing physical behavior. Selected 2D metamaterials consisting
of graphitic nanostructures, sometimes called ``nanoporous
graphene'', have been
synthesized~\cite{{treier2011},{moreno18},{Porgra09}} and could
find their use in hydrogen storage, gas separation and
purification, DNA sequencing, and as magnets and
supercapacitors~\cite{{xu2012},{shengbai2016}}. One of such
graphitic nanostructures is the phenalenyl radical, which is
obtained by single deprotonation of the phenalene molecule.
Similar to many other related and extensively studied
structures~\cite{{Nakano07},{Nakano10},{Nakano05}}, this radical
is stable and displays a net magnetic moment of
$1$~$\mu_{\rm{B}}$~\cite{{Sogo1957},{Boer1956},{McConnell1957},%
{McConnell1958},{Reid1958},{Gerson1966},{Haddon1978},{Kazuyuki2016}}.
Phenalenyl radicals may polymerize and form a 2D ``nanoporous
graphene'' metamaterial that may be deformed by strain to display
a complex behavior of the Poisson's ratio including strain-related
changes in value and sign~\cite{DT271}. Strain-related changes in
coupling between connected phenalenyl radicals, which carry a
nonvanishing magnetic moment, may change not only the electronic
structure of the system, but also its magnetic order. Elemental
substitution of carbon by boron or nitrogen in the radicals may
cause additional drastic changes in the magnetic behavior.

Here, we report the design of a Kagom\'{e}-like 2D metamaterial,
which is composed of phenalenyl radicals covalently interconnected
at their corners. This pristine metamaterial exhibits a strong
ferromagnetic order, which can be tuned significantly by applied
strain at little energy cost. In this metamaterial, externally
applied strain primarily turns the phenalenyl radicals with
respect to each other. We find that this deformation, which
changes the magnetic moment of individual radicals, is accompanied
by a charge flow that may change the system from a metal to a
narrow-gap semiconductor. Substitutional doping of this
metamaterial, caused by replacing carbon by either boron or
nitrogen atoms, decreases the net magnetization significantly.
However, simultaneous doping by B and N changes the magnetic order
to antiferromagnetic. Unlike in the pristine system,
strain-induced turning of the doped phenalenyl radicals does not
change their magnetic moment. The structures described here are
constrained to planar geometries during deformation, which can be
realized by being attracted to a planar substrate or in a sandwich
geometry.

\begin{figure}[h]
\includegraphics[width=1.0\columnwidth]{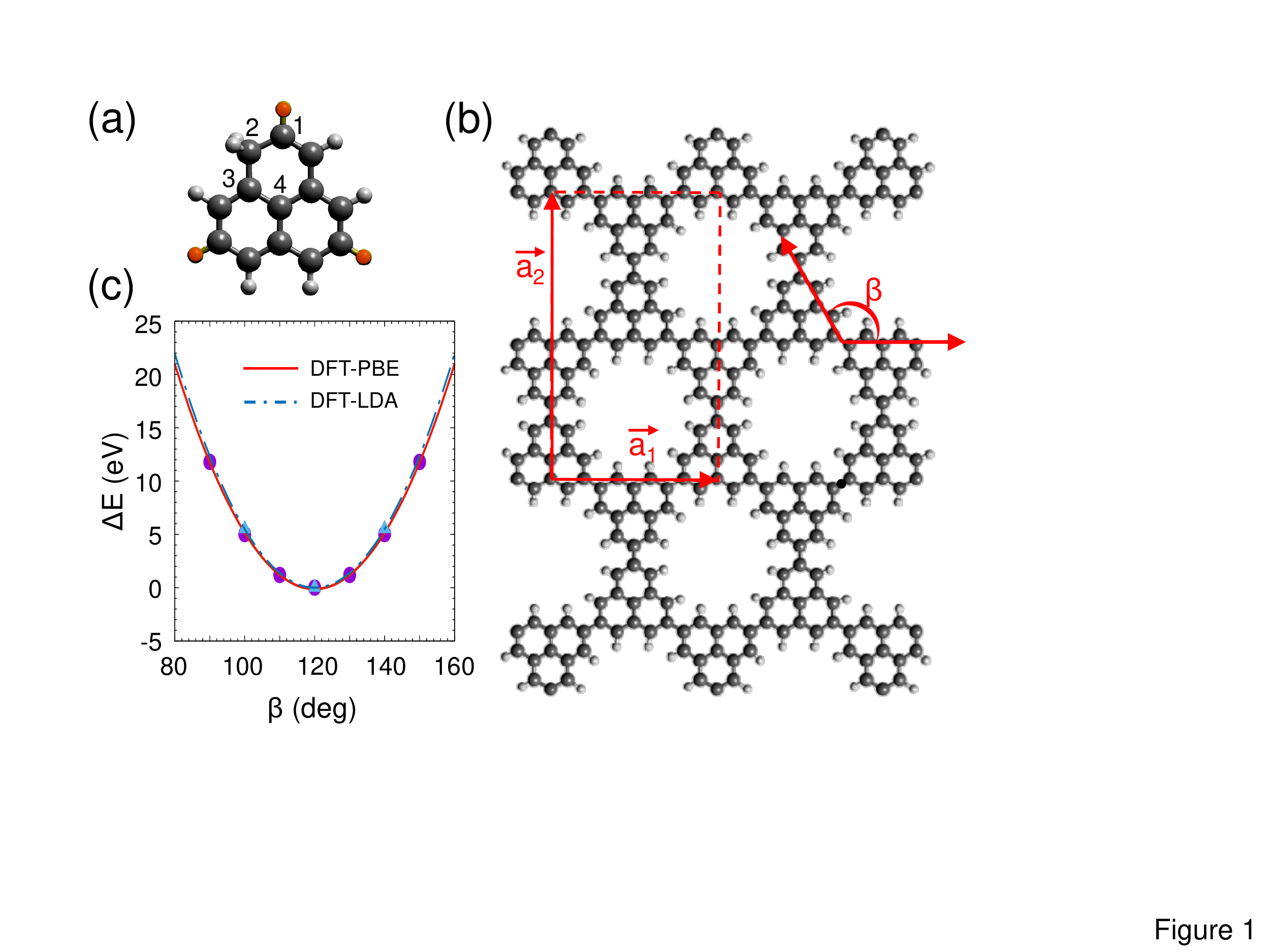}
\caption{%
Polymerization of phenalenyl radical units into a Kagom\'{e}-like
2D polyphenalenyl lattice. %
(a) Phenalene (1$H$-Phenalene, C$_{13}$H$_{10}$), polycyclic
    aromatic hydrocarbon (PAH), used to produce the phenalenyl
    radical by deprotonation of one of the corner atoms.
(b) (C$_{13}$H$_6$)$_\infty$ polyphenalenyl lattice. %
    The primitive unit cell is delimited by the lattice vectors
    $\vec{a}_1$ and $\vec{a}_2$. The three hydrogens
    in phenalene, highlighted in (a), detach when forming polyphenalenyl. %
(c) Total energy change ${\Delta}E$ per unit cell caused by the
    changing orientation angle $\beta$, obtained using DFT-PBE
    (red solid line) and DFT-LDA (blue dashed line).
\label{fig1}}
\end{figure}

\section{Results}

\subsection{Kagom\'{e}-like 2D polyphenalenyl as magnetic metamaterial}

The optimized phenalenyl radical, also called perinaphthenyl,
obtained by mono-deprotonation of the phenalene molecule and shown
in Fig.~\ref{fig1}(a), is composed of 13 carbon and 9 hydrogen
atoms (${\rm C}_{13}{\rm H}_9$). Since the phenalenyl radical is
missing one H compared to the non-magnetic phenalene, it carries
one unpaired electron, resulting in a net magnetic moment of
1~$\mu_{\rm{B}}$. Within the H\"{u}ckel molecular approximation,
the non-bonding singly-occupied molecular orbital (SOMO) of ${\rm
C}_{13}{\rm H}_9$ locates the spin density in a uniform and
regular fashion on every other of the 12 equivalent peripherical
conjugated carbon atoms~\cite{trinquier2017}. The macroscale 2D
metamaterial polyphenalenyl, shown in Fig.~\ref{fig1}(b), is an
optimized structure of polymerized phenalenyl radicals that are
covalently connected at the corners. There are two identical
phenalenyl radicals with different orientation in the primitive
unit cell of polyphenalenyl, as seen in Fig.~\ref{fig1}(b). To
provide for increased freedom in the magnetic order, we choose a
unit cell that is rectangular and twice as large as the primitive
unit cell.

To save computer time when considering many 2D polyphenalenyl
metamaterials with different relative orientation angles $\beta$,
we first optimized the connection distance between frozen
phenalenyl radicals and then relaxed the hydrogen atoms while
keeping the carbon atoms frozen. We refer to this optimization
strategy as `partial global relaxation' in the discussion
hereafter. Compared to a more accurate and time-consuming global
relaxation, structures optimized in this way were found to be less
stable by only ${\lesssim}21$~meV/C-atom with respect to globally
optimized structures. We found also only small differences of
${\lesssim}0.6$\% between structures optimized globally or in the
simplified way. We concluded that the geometry and stability of
globally optimized structures are reproduced very well by partial
global relaxation and used the latter approach to optimize all
geometries in this study unless stated otherwise. We find
energetic preference for ferromagnetic ordering in the optimized
2D metamaterial. The calculated magnetic moment of one unit cell,
which contains four radicals, is ${\approx}3.6$~$\mu_{\rm{B}}$. We
found all phenalenyl radicals in the unit cell to carry the same
magnetic moment of 0.9~$\mu_{\rm{B}}$, which is somewhat smaller
than the 1~$\mu_{\rm{B}}$ value of an isolated phenalenyl radical
and is a consequence of polymerization.

\begin{figure}[t]
\includegraphics[width=1.0\columnwidth]{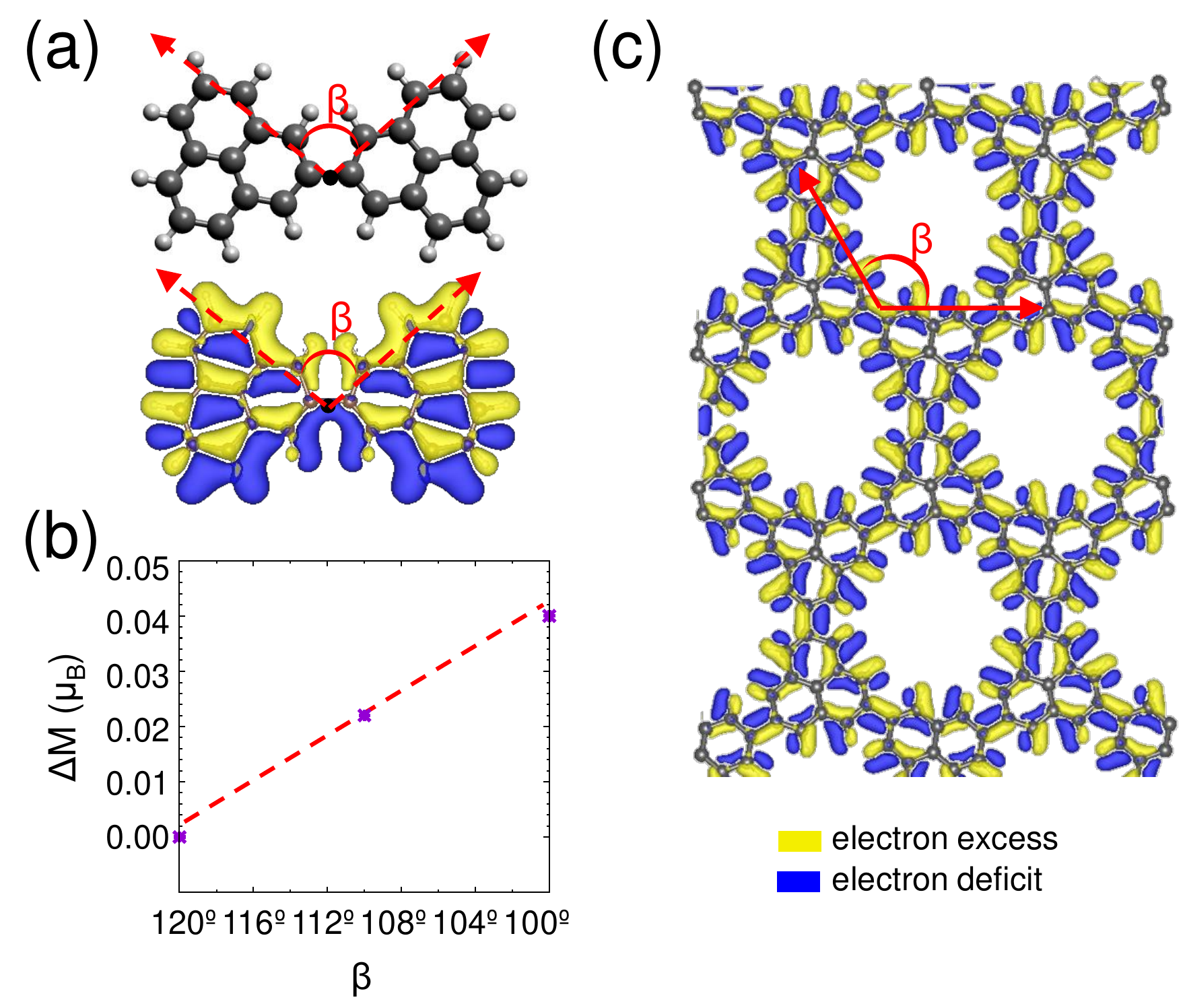}
\caption{%
(a) Top view of the deformed phenalenyl dimer C$_{26}$H$_{18}$
    with $\beta=100^\circ$ in the upper panel. Charge
    redistribution in the phenalenyl dimer caused by deviation
    from the equilibrium structure with $\beta=120^\circ$ is shown
    in the lower panel, which depicts isosurface bounding regions
    of electron excess at $+3.2{\times}10^{-2}$~{e}/{\AA}$^3$
    (yellow) and electron deficiency at
    $-3.2{\times}10^{-2}$~{e}/{\AA}$^3$ (blue). %
(b) Difference between the relative contribution of the upper and
    lower atoms towards the net magnetic moment $M=2~\mu_{\rm{B}}$,
    presented as a function of the relative orientation angle $\beta$
    in the phenalenyl dimer C$_{26}$H$_{18}$. %
(c) Charge redistribution in strained polyphenalenyl with
    $\beta=100^\circ$, caused by deviation from the equilibrium
    structure with $\beta=120^\circ$. %
\label{fig2}}
\end{figure}

Under compressive strain, deformed structures can be characterized
by the angle $\beta$, defined in Fig.~\ref{fig1}(b), which
describes the relative orientation of adjacent radicals.
$\beta=120^\circ$ in the optimized structure of 2D polyphenalenyl.
$\beta$ may increase or decrease under in-plane compressive
strain, but will not change under tension. The deformation energy
${\Delta}E$ depends on the deformation characterized by the angle
$\beta$. As seen in Fig.~\ref{fig1}(c), ${\Delta}E$ exhibits a
harmonic change around the equilibrium value ${\beta}=120^\circ$.
We find that changing $\beta$ by $20^\circ$ requires $5$~eV per
unit cell or $96$~meV/C-atom, which is only ${\approx}2$\% of the
bond strength. This means this 2D polyphenalenyl is a soft
metamaterial that can be deformed easily by applying in-plane
stress.

Changing $\beta$ by in-plane compression causes a net charge flow
in the structure. We first study the charge flow in a dimer made
of two phenalenyl radicals connected by one C--C bond, shown in
Fig.~\ref{fig2}(a) for ${\beta}=100^\circ$. Decreasing $\beta$
from the equilibrium value, we find that more electrons accumulate
at the edges that are approaching during the deformation, and
deplete at the edges that are getting more distant. Even though
the net charge and net magnetic moment of $2$~$\mu_{\rm{B}}$ of
the system remain the same, the local charge density and the local
magnetic moment may change as $beta$ changes. We characterize the
difference between the sum of local magnetic moments in atoms on
the upper and lower edges of one of the radicals by the quantity
${\Delta}M$. As we show in Fig.~\ref{fig2}(b), ${\Delta}M$
increases monotonically along with charge density differences as
$\beta$ deviates from the equilibrium value of $120^\circ$.

Unlike in the dimer discussed above, each phenalenyl radical is
connected not to one, but three radicals in 2D polyphenalenyl, as
seen in Fig.~\ref{fig2}(c). As individual radicals rotate in an
alternate way during deformation, each edge of any radical
approaches and distances itself from edges of two adjacent
radicals, compensating for the charge flow from the two sides.
There is no net charge accumulation on any of the edges, but still
a local charge redistribution within each radical. This changes
the electronic structure and magnetic behavior of the 2D
polyphenalenyl system.

\begin{figure}
\includegraphics[width=1.0\columnwidth]{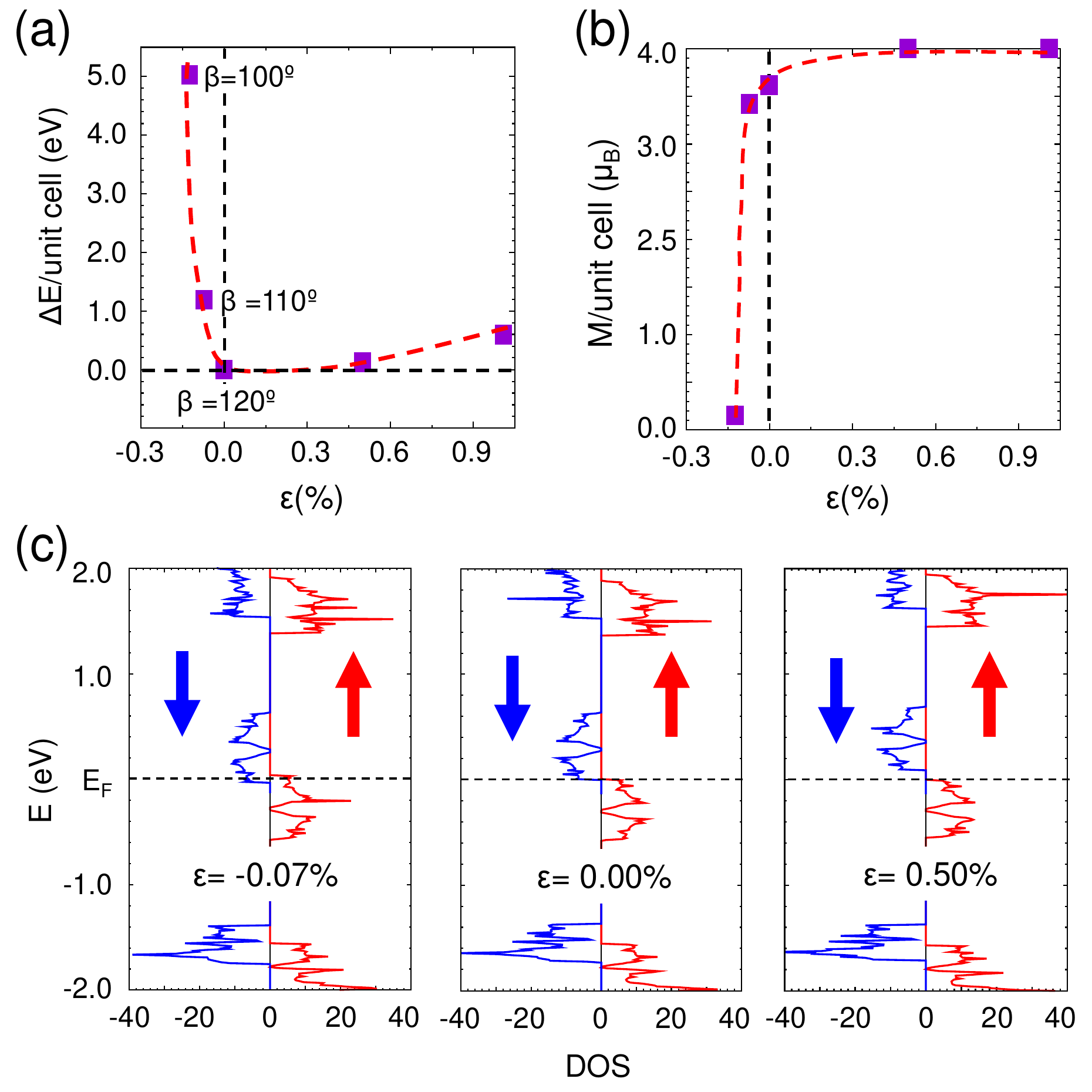}
\caption{%
Strain-dependent properties of the Kagom\'{e}-like 2D
polyphenalenyl metamaterial. %
(a) Deformation energy ${\Delta}E$ as a function of in-plan strain
$\epsilon$. %
(b) Total magnetic moment $M$ per unit cell as a function of
$\epsilon$. %
(c) Spin-polarized density of states (DOS) for different values of
$\epsilon$. The different spin polarizations are distinguished by
color.
\label{fig3}}
\end{figure}

\subsection{Strain-controlled electronic and magnetic properties
of the Kagom\'{e}-like 2D polyphenalenyl metamaterial}

As mentioned above, deformation of the 2D polyphenalenyl structure
under in-plane compressive strain can be characterized by changes
in the orientation angle $\beta$. Under uniform in-plane tensile
strain, ${\beta}=120^\circ$ does not change and there is only an
increase in the C--C bond length. The deformation energy as a
function of in-plane strain is shown in Fig.~\ref{fig3}(a).
Results provided for the ${\epsilon}<0$ range in this figure
correspond to those for ${\beta}$ deviating from the $120^\circ$
equilibrium value in Fig.~\ref{fig1}(c). Our results for
${\epsilon}>0$ indicate that the systems is rather soft also under
tension, since 1\% tensile strain requires an energy investment of
only ${\Delta}E<1$~eV per unit cell.

Lattice deformation also affects the electronic and thus the
magnetic structure of the system. As seen in Fig.~\ref{fig3}(b),
the magnetic moment per unit cell $M$ is particularly sensitive to
compressive strain that changes the angle $\beta$. In comparison
to $M=3.6$~$\mu_{\rm{B}}$ in the unstrained lattice, the magnetic
moment reduction to $M=0.1$~$\mu_{\rm{B}}$ under
${\epsilon}=-0.12$\% compression is significant. We find the
magnetic moment to increase less rapidly under tension, achieving
$M=4.0$~$\mu_{\rm{B}}$ at ${\epsilon}=+1.0$\% due to reduced
coupling between the phenalenyl radicals.

The magnetic behavior of the system under strain is shown in the
spin-polarized DOS in Fig.~\ref{fig3}(c). There is a majority band
(spin-up) that overlaps partly with the minority band (spin-down),
rendering the system metallic under zero or compressive strain. It
is this increasing band overlap that is responsible for quenching
the magnetic moment under increasing compressive strain.
Increasing tensile strain beyond ${\epsilon}{\gtrsim}0.5$\% opens
a narrow semiconducting gap between the majority and minority
bands and the magnetic moment saturates at $M=4.0$~$\mu_{\rm{B}}$.
The system remains ferromagnetic in the entire deformation range
discussed here.

\begin{figure}[t]
\includegraphics[width=1.0\columnwidth]{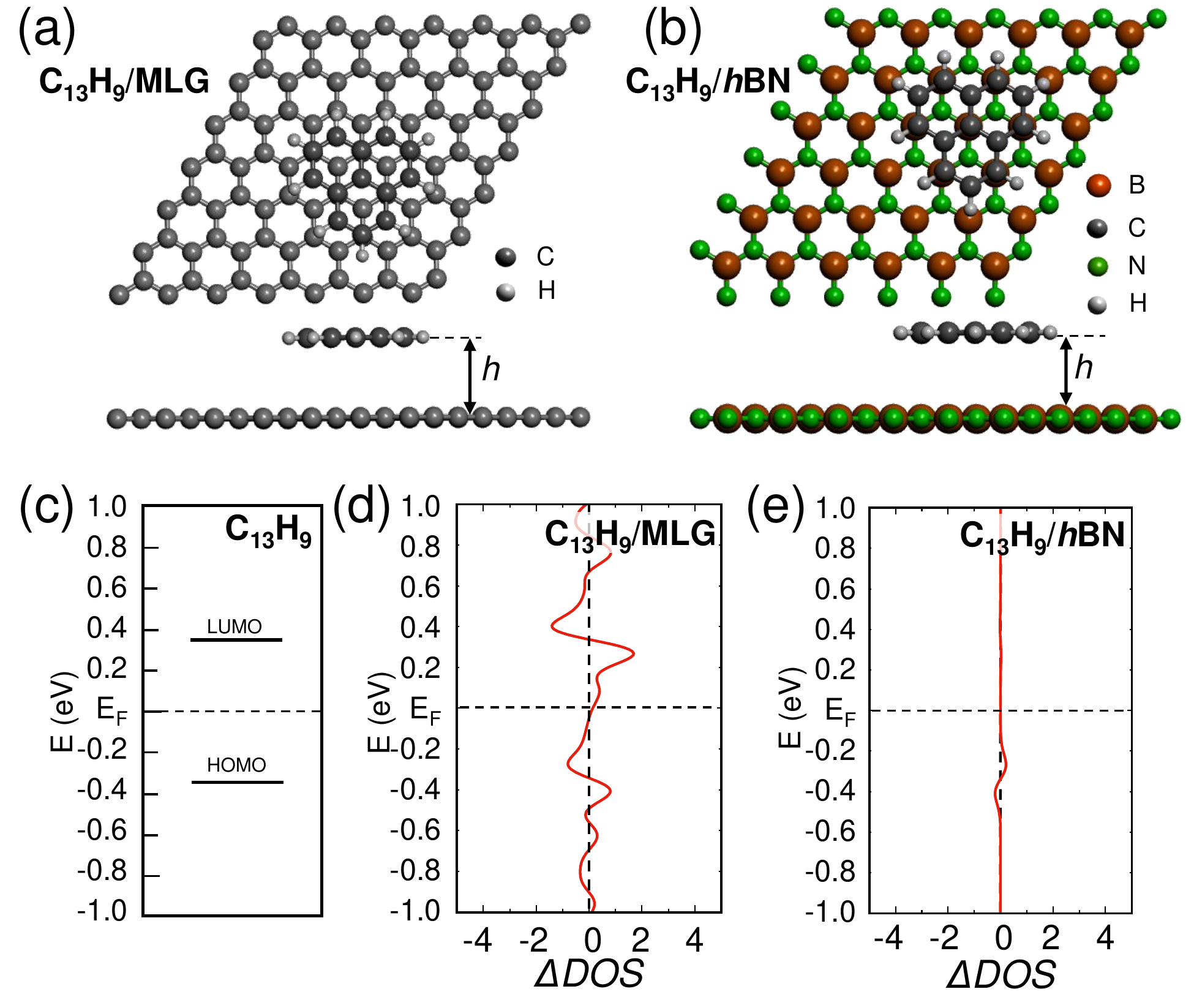}
\caption{%
Equilibrium structure and adsorption-induced changes in the DOS of
C$_{13}$H$_9$ adsorbed on graphene and $h-$BN. %
Top and side views of the adsorption geometry of C$_{13}$H$_9$ on
(a) graphene and (b) $h-$BN substrates. %
(c) The DFT energy level spectrum of a C$_{13}$H$_9$ molecule. %
Adsorption-induced changes in the DOS of phenalenyl on (d)
graphene and (e) $h-$BN, with
${\Delta}DOS=DOS({\rm{C}}_{13}{\rm{H}}_9/{\rm{substrate}})
-DOS({\rm{C}}_{13}{\rm{H}}_9)-DOS({\rm{substrate}})$. %
All results are based on superlattice calculations.
\label{fig4}}
\end{figure}

\subsection{Adsorption of phenalenyl radicals on graphene and $h-$BN}

Clearly, the Kagom\'{e}-like 2D polyphenalenyl system needs to be
stabilized by a substrate against structural collapse. An ideal
substrate should be stable and capable of supporting the
Kagom\'{e}-like 2D lattice, but should not affect its electronic
and magnetic structure. We have considered monolayers of graphene
and hexagonal boron nitride ($h-$BN) as suitable substrates that
are both stable and chemically non-reactive.

The adsorption geometry of C$_{13}$H$_9$ radicals on graphene and
$h-$BN is shown for one unit cell of a superlattice in
Fig.~\ref{fig4}(a-b). We find the equilibrium separation between
phenalenyl and the substrate to be $h=4.0$~{\AA} for graphene and
$h=3.9$~{\AA} for $h-$BN. These separations are similar to the
interlayer distance of graphite and $h-$BN.

The DFT-based energy level spectrum of an isolated C$_{13}$H$_9$
radical is presented in Fig.~\ref{fig4}(c). We expect the weak
interaction between C$_{13}$H$_9$ and the substrate not to affect
its electronic structure much. To quantify this effect, we present
${\Delta}$DOS as the difference between the DOS of the combined
system and the superposition of densities of states of the
isolated radicals %
and the isolated substrate in Fig.~\ref{fig4}(d) for graphene and
Fig.~\ref{fig4}(e) for $h-$BN. As expected, we find ${\Delta}$DOS
very small in comparison to the DOS of polyphenalenyl in
Fig.~\ref{fig3}(c). Consequently, the effect of
adsorbate-substrate interaction on the electronic and related
magnetic structure of phenalenyl can be neglected. In that case,
results for a free-standing Kagom\'{e}-like 2D polyphenalenyl
system should apply also in presence of substrates such as
graphene and $h-$BN, or in a sandwich geometry between such layers. %
Preventing polymerization due to collapse of the 2D polyphenalenyl
monolayer or by allowing multilayer formation is a key to
maintaining
the intriguing magnetic behavior described here. %

\begin{figure*}[t]
\includegraphics[width=1.8\columnwidth]{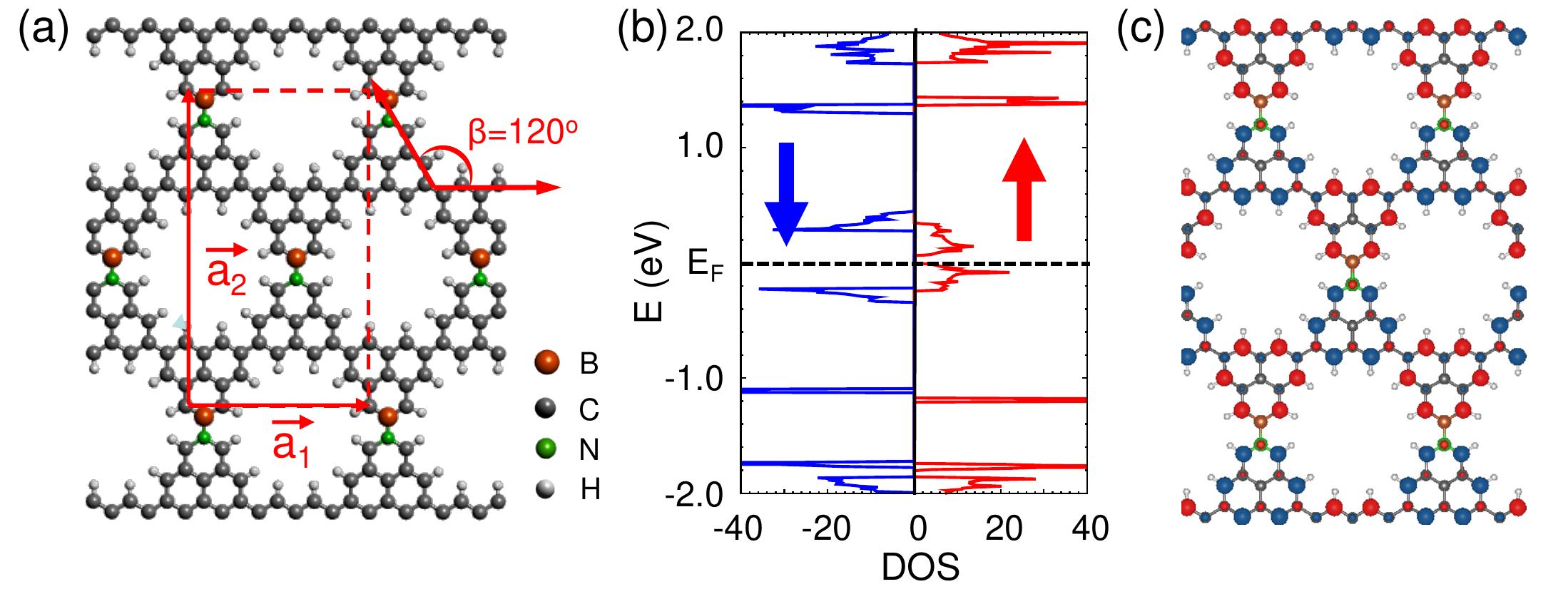}
\caption{%
Kagom\'{e}-like (C$_{12}$H$_6$B-C$_{12}$H$_6$N)$_\infty$ 2D
polyphenalenyl lattice doped with B and N atoms. %
(a) Top view of the equilibrium lattice structure with
    $\beta=120^\circ$. The conventional unit cell,
    delimited by the $\vec{a}_1$ and $\vec{a}_2$
    basis vectors, contains two formula units. %
(b) Spin-polarized DOS of the doped polyphenalenyl lattice
    shown in panel (b). Spin-up and spin-down DOS are distinguished
    by color. %
(c) Spin polarized charge density $\rho_\uparrow-\rho_\downarrow$
    in the B-N doped 2D polyphenalenyl lattice of panel (a).
    The isosurfaces are bounded by %
    $+2{\times}10^{-3}$~e/{\AA}$^3$ for the
      $\uparrow$ spin (red) and by %
    $-2{\times}10^{-3}$~e/{\AA}$^3$ for the
      $\downarrow$ spin (dark blue). %
\label{fig5}}
\end{figure*}

\subsection{Effect of doping on magnetic properties of strained
Kagom\'{e}-like 2D polyphenalenyl}

As mentioned above, the phenalenyl radical ${\rm C}_{13}{\rm
H}_{9}$ carries the magnetic moment $1$~$\mu_{B}$ due to one
unpaired electron in the radical. Replacing a carbon atom in the
phenalenyl radical by boron with one less valence electron or by
nitrogen with one more valence electron will change the charge of
the radical by one, allowing all electrons to pair. Consequently,
doped phenalenyl radicals C$_{12}$BH$_{9}$ and C$_{12}$NH$_{9}$
should be non-magnetic. In the phenalenyl radical displayed in
Fig.~\ref{fig1}(a), we can distinguish four inequivalent carbon
sites, labeled $1-4$, among the 13 carbon atoms in total.
Replacing any of these 4 C atoms by B or N quenched the magnetic
moment of the doped radical to zero as suggested above.

Optimum location of the dopant atoms within the 2D metamaterial is
driven by stability considerations. As a guiding principle, we
compared the bond strength of C--C, B--N, C--B and C--N dimers.
Among these, the C--C bond was most stable, followed by the 1.9~eV
less stable B--N bond. The C--N bond was found to be almost as
stable as the B--N bond, and the C--B bond was found to be 2.8~eV
less stable than the B--N bond.
Based on these findings, a stable 2D metamaterial allotrope may
consist of phenalenyl radical pairs connected by B--N instead of
C--C bonds, as seen in Fig.~\ref{fig5}(a).

The (C$_{12}$H$_6$B-C$_{12}$H$_6$N)$_\infty$ 2D polyphenalenyl
lattice doped with B and N atoms, shown in Fig.~\ref{fig5}(a),
carries zero magnetization, as illustrated by the symmetric
spin-polarized DOS in Fig.~\ref{fig5}(b). In contrast to the
pristine metamaterial which is a metal, the doped metamaterial is
a semiconductor with a narrow band gap of about $0.06$~eV seen in
Fig.~\ref{fig5}(b).

The spin-polarized charge density $\rho_\uparrow-\rho_\downarrow$
of the doped polyphenalenyl lattice is shown in
Fig.~\ref{fig5}(c), with the two spin polarizations being
distinguished by color. We clearly see that the dominant color on
the B-doped phenalenyl radical is red, indicating a majority state
with spin-up electrons and a net magnetic moment of
${\approx}0.35$~$\mu_{\rm{B}}$. The dominant color on the N-doped
phenalenyl radical is dark blue, indicating that the majority
state carries spin-down electrons and a net magnetic moment of
${\approx}-0.35$~$\mu_{\rm{B}}$. This is somewhat unexpected,
since separated doped radicals were found to be non-magnetic, and
indicates the role of coupling and charge transfer between
connected radicals. With magnetic moments of opposite direction on
adjacent phenalenyl radicals in the unit cell, the 2D metamaterial
is antiferromagnetic.

Same as in the pristine metamaterial, in-plane compressive strain
changes the orientational angle $\beta$ of the doped phenalenyl
radicals. In B--N doped polyphenalenyl, changing ${\beta}$ by
$20^\circ$ from the equilibrium value requires an energy
investment of $4.46$~eV/unit cell, slightly less than in the
pristine system according to Fig.~\ref{fig3}(a). As we showed in
Fig.~\ref{fig3}(b), even minute compressive strain that changed
${\beta}$ by $20^\circ$ caused a significant reduction of the
magnetic moment in the pristine system. In the B--N doped system,
on the other hand, the same change of ${\beta}$ by $20^\circ$ has
very little effect on the local magnetization of B- and N-doped
phenalenyl radicals. Even under in-plane compressive strain, the
doped 2D polyphenalenyl remains a narrow-gap semiconductor and
anti-ferromagnet. Additional results depicting changes in the
spin-polarized charge density of pristine and doped polyphenalenyl
caused by changes in ${\beta}$ are presented in the Supporting
Material~\cite{SM-2DMM19}.

The 2D metamaterial composed of C$_{13}$H$_{9}$ phenalenyl
radicals carrying a $1$~$\mu_{\rm{B}}$ magnetic moment is just one
member of a large family of polycyclic aromatic hydrocarbons
(PAHs). Other PAH molecules or radicals have different magnetic
properties, including the non-magnetic C$_{14}$H$_{10}$
phenanthrene molecule considered as component of a 'mechanical'
metamaterial described in Ref.~\cite{DT271}. When
substitutionally doped with B or N, also this molecule should
carry a $1$~$\mu_{\rm{B}}$ magnetic moment according to Hund's
rule. Consequently, a 2D metamaterial consisting of polymerized
pristine C$_{14}$H$_{10}$ molecules should be non-magnetic. Doping
every other molecule by boron and the remaining molecules by
nitrogen, polyphenanthrene may acquire interesting magnetic
behavior. The effect of B or N doping in polyphenalenyl and
polyphenanthrene is thus opposite. With specific applications in
mind, other PAH molecules can be chosen as building blocks of 2D
metamaterials.

We have seen that doping every other phenalenyl radical in
polyphenalenyl by B and the remaining radicals by N has changed
the magnetic order from ferromagnetic in the pristine system to
antiferromagnetic in the doped metamaterial. The cause underlying
this change was the charge transfer in the polar B--N links
connecting the radicals. There are other methods to modify the
charge distribution in 2D systems including a `van der Waals'
2D/2D contact with a substance like the Ca$_{2}$N electride, which
has been shown to transfer a significant charge to systems such as
2D boron~\cite{DT272}. In the rigid-band model, electron transfer
to B--N doped polyphenalenyl would raise the Fermi level and fill
initially empty spin-up and spin-down bands to a different degree,
leading to a nonzero global magnetization. Since B-doped and
N-doped phenalenyl radicals carry opposite net charges causing
opposite magnetic moments, uniform electron doping would break
this symmetry in charge doping and magnetic moments, turning the
system ferrimagnetic.

\section{Summary and Conclusions}

We have used {\em ab initio} spin-polarized DFT calculations to
study the magnetic order in a Kagom\'{e}-like 2D metamaterial
consisting of pristine or substitutionally doped phenalenyl
radicals polymerized into a nanoporous, graphene-like structure.
In this and in a larger class of related structures, the
constituent PAH molecules can be considered as quantum dots that
may carry a net magnetic moment. The structure of this porous
system is rather soft and may be changed at little energy cost by
applying in-layer strain. Structural changes modify the coupling
between such quantum dots, causing a change in the electronic and
magnetic structure. The pristine polyphenalenyl material is
ferromagnetic, but its magnetization may be changed by a factor of
two by applying moderate strain. Doping every other radical by B
or N atoms turns the system antiferromagnetic, with local magnetic
moments rather independent of strain-related structural changes.
We believe this is only one interesting example of how to control
the magnetic order and the underlying electronic structure in
magnetic metamaterials.

\section{Computational Techniques}

We have studied the electronic and magnetic properties as well as
the deformation energy of polyphenalenyl using {\em ab initio}
density functional theory (DFT) as implemented in the
\textsc{VASP} code~\cite{{VASP1},{VASP2},{VASP3}}. We represented
this 2D structure by imposing periodic boundary conditions in all
directions and separating individual layers by a vacuum region of
$15$~{\AA}. We used projector-augmented-wave (PAW)
pseudopotentials~\cite{{PAW1},{PAW2}} and the
Perdew-Burke-Ernzerhof (PBE)~\cite{PBE} or the Local Density
Approximation (LDA)~\cite{{Ceperley1980},{Perdew81}}
exchange-correlation functionals. The Brillouin zone of the
conventional unit cell of the 2D structure was sampled by an
$11{\times}5{\times}1$ $k$-point grid~\cite{Monkhorst-Pack76}. We
used a value of $500$~eV as the electronic kinetic energy cutoff
for the plane-wave basis and a total energy difference between
subsequent self-consistency iterations below $10^{-4}$~eV as the
criterion for reaching self-consistency. All geometries have been
optimized using the conjugate-gradient method~\cite{CGmethod},
until none of the residual Hellmann-Feynman forces exceeded
$10^{-2}$~eV/{\AA}.


\section*{Acknowledgments}

D.L. and D.T. acknowledge financial support by the NSF/AFOSR EFRI
2-DARE grant number EFMA-1433459. Computational resources have
been provided by the Michigan State University High Performance
Computing Center. Sandia National Laboratories is a multi-mission
laboratory managed and operated by National Technology and
Engineering Solutions of Sandia, LLC., a wholly owned subsidiary
of Honeywell International, Inc., for the U.S. Department of
Energy's National Nuclear Security Administration under Contract
DE-NA0003525. The views expressed in the article do not
necessarily represent the views of the U.S. DOE or the United
States Government.


\end{document}


\setcounter{subsection}{0} %
\renewcommand\thesubsection{\Alph{subsection}}
\setcounter{equation}{0}
\renewcommand{\theequation}{S\arabic{equation}}
\setcounter{figure}{0}
\renewcommand{\thefigure}{S\arabic{figure}}

\subsubsection{Spin polarized charge density}

\begin{figure}[t]
\includegraphics[width=1.0\columnwidth]{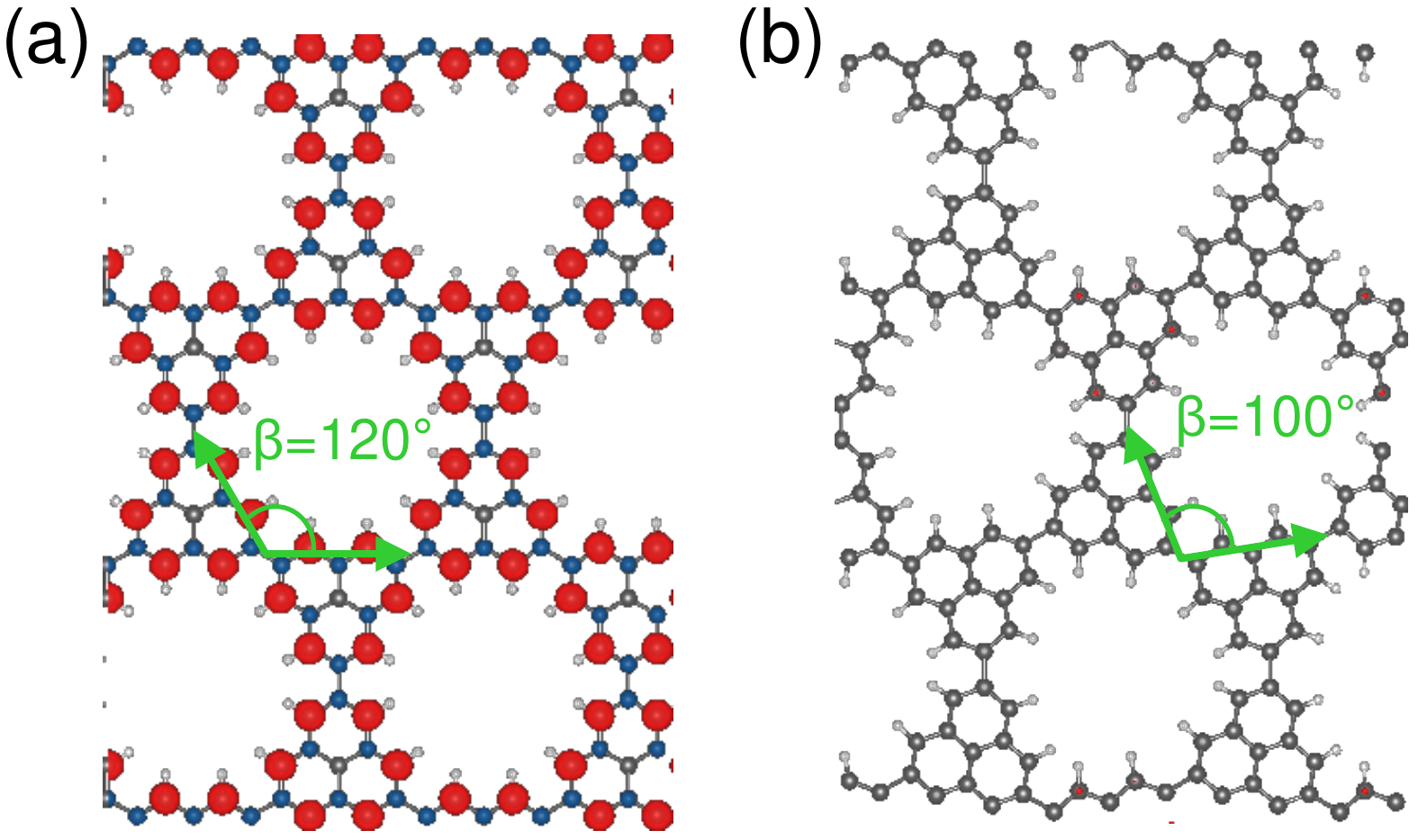}
\caption{%
Spin polarized charge density $\rho_\uparrow-\rho_\downarrow$ in
the (C$_{13}$H$_6$-C$_{13}$H$_6$)$_\infty$ polyphenalenyl
lattice, with isosurfaces bounded by %
$+3{\times}10^{-3}$~e/{\AA}$^3$ for the $\uparrow$ majority spin
(red) and by %
$-3{\times}10^{-3}$~e/{\AA}$^3$ for the $\downarrow$ minority spin
(dark blue). %
The two values of the opening angle $\beta$ shown are
(a) $\beta=120^\circ$ and %
(b) $\beta=100^\circ$. %
\label{figS1}}
\end{figure}

The spin-polarized charge density $\rho_\uparrow-\rho_\downarrow$
of the (C$_{13}$H$_6$-C$_{13}$H$_6$)$_\infty$ polyphenalenyl
lattice with two values of the orientational angle $\beta$ is
shown in Fig.~\ref{figS1}. With the two spin polarizations being
represented by different colors, we clearly see that the majority
spin, represented by red, dominates the lattice, indicating that
the (C$_{13}$H$_6$-C$_{13}$H$_6$)$_\infty$ system is
ferromagnetic.

\begin{figure}[t]
\includegraphics[width=1.0\columnwidth]{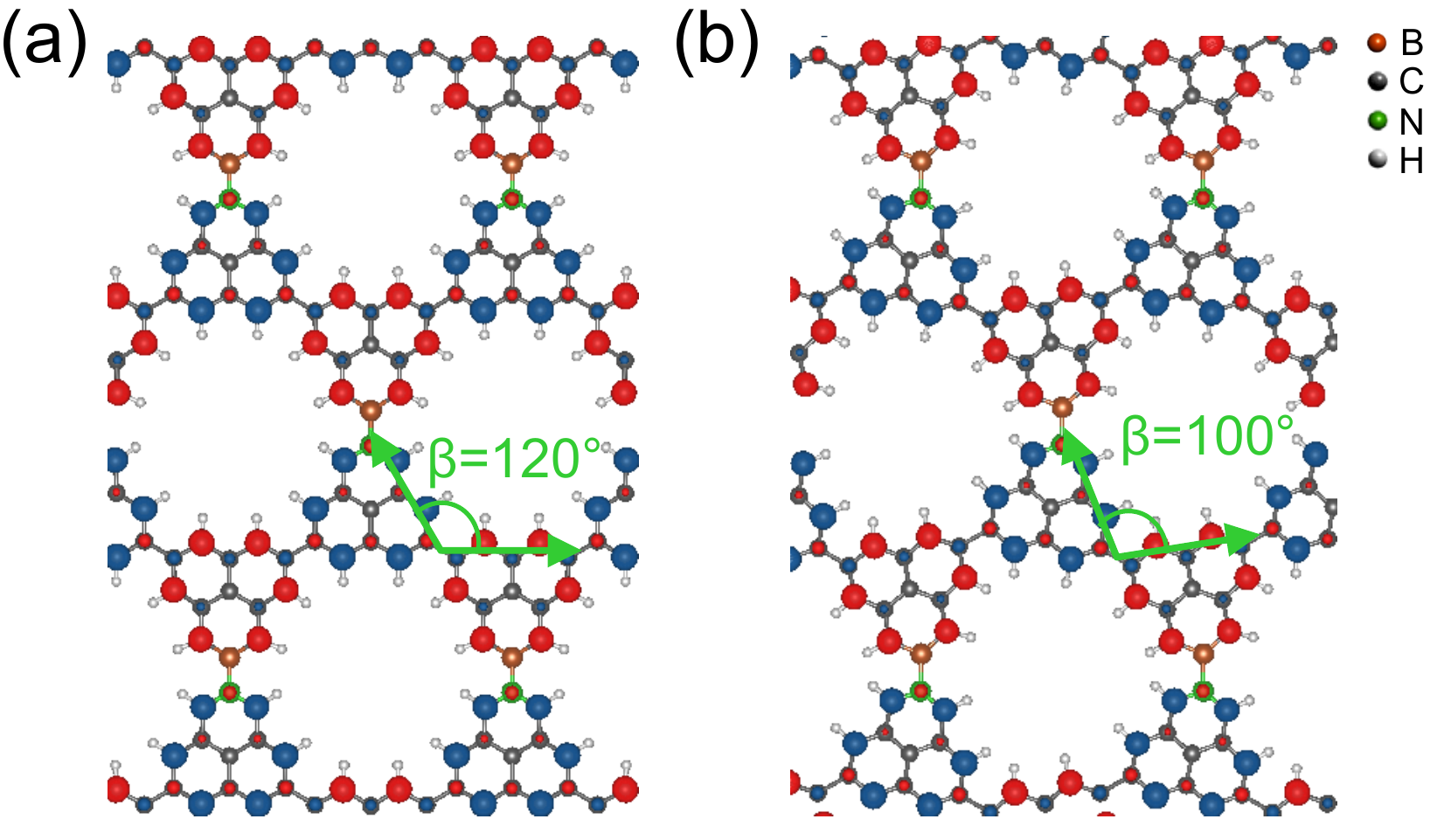}
\caption{%
Spin polarized charge density $\rho_\uparrow-\rho_\downarrow$ in
the B-N doped (C$_{12}$H$_6$B-C$_{12}$H$_6$N)$_\infty$
polyphenalenyl lattice, with isosurfaces bounded by %
$+2{\times}10^{-3}$~e/{\AA}$^3$ for the $\uparrow$ spin
(red) and by %
$-2{\times}10^{-3}$~e/{\AA}$^3$ for the $\downarrow$ spin
(dark blue). %
The two values of the opening angle $\beta$ shown are
(a) $\beta=120^\circ$ and %
(b) $\beta=100^\circ$. %
\label{figS2}}
\end{figure}

The spin-polarized charge density of the B--N doped
(C$_{12}$H$_6$B-C$_{12}$H$_6$N)$_\infty$ polyphenalenyl lattice
with two values of the orientational angle $\beta$ is shown in
Fig.~\ref{figS2}. The two spin polarizations are distinguished by
color. We clearly see the dominant color on the B-doped radical to
be red, indicating that the majority state carries spin-up
electrons. The dominant color on the N-doped radical is dark blue,
indicating that the majority state carries spin-down electrons.
With magnetic moments of opposite direction on adjacent phenalenyl
radicals in the unit cell, the system is antiferromagnetic.

Comparing results for both systems for different values of the
orientation angle $\beta$, we clearly see that the magnetization
is very sensitive to $\beta$ in the undoped system, as seen by
comparing Figs.~\ref{figS1}(a) and \ref{figS1}(b). The magnetic
moment of the two sublattices in the doped system, represented in
Fig.~\ref{figS2}, is much less sensitive to $\beta$ due to the
charge depletion on the polar B--N bonds connecting the
sublattices.
